# Robust acoustic pulling using chiral surface waves

Neng Wang[1,*], Ruo-Yang Zhang[2], and C. T. Chan[2]

[1]*Institute of Microscale Optoelectronics, Shenzhen University, Shenzhen 518060, China*

[2]*Department of Physics, The Hong Kong University of Science and Technology, Clear Water Bay, Hong Kong, China*

*Corresponding to: [nwang17@szu.edu.cn](mailto:nwang17@szu.edu.cn)*

**Abstract**:

We show that long-range and robust acoustic pulling can be achieved by using a pair of one-way chiral surface waves supported on the interface between two phononic crystals composed of spinning cylinders with equal but opposite spinning velocities embedded in water. When the chiral surface mode with a relative small Bloch wave vector is excited, the particle located in the interface waveguide will scatter the excited surface mode to another chiral surface mode with a greater Bloch wave vector, resulting in an acoustic pulling force, irrespective of the size and material of the particle. Thanks to the backscattering immunity of the chiral surface waves against local disorders, the particle can be pulled following a flexible trajectory as determined by the shape of the interface. As such, this new acoustic pulling scheme overcomes some of the limitations of the traditional acoustic pulling using structured beams, such as short pulling distances, straight-line type pulling and strong dependence on the scattering properties of the particle. Our work may also inspire the application of topological acoustics to acoustic manipulations.

# I. Introduction

The ability of acoustic waves to exert radiation forces and torques on matter enables contactless and noninvasive acoustic manipulations, which have found fruitful applications in various areas ranging from physics [1-4], to chemistry[5, 6] and biology [7-10]. Among these manipulations, the acoustic pulling [11-18], which refers to pulling particles using the acoustic waves towards the source, is perhaps most amazing. In addition to being a counter-intuitive phenomenon, the acoustic pulling also provides a new mechanism for the acoustic manipulations apart from the levitation [19-22], trapping [23-26] and binding [27, 28].

The acoustic pulling force stems from the backward momentum gained by the particle from the acoustic wave during the scattering process. Due to the momentum conservation law, this requires that the scattered wave has a greater forward momentum than the incident one. To achieve pulling in free space, structured beams such as a Bessel beam with a very large cone angle [11-15] or multiple beams with giant obliquely incident angles [16, 17], are required to minimize the forward momentum of the incident wave. The technical difficulty on realizing long-range Bessel beams and the finite beam waists of the obliquely incident beams limits the application scenarios of acoustic pulling in reality. The size, shape and material of the particle also need to be carefully customized to enhance the forward scattering and reduce the backward scattering simultaneously so that the forward momentum of the scattered wave can be large enough. Therefore, the free space acoustic pulling is not only limited to short-range in general but also dependent strongly on the scattering properties of the manipulated object.

In this paper, we propose a way to achieve long-range and robust acoustic pulling for arbitrary particles located in an interface waveguide sandwiched between two phononic crystals composed of oppositely spinning cylinders embedded in water. On the interface of the two phononic crystals, there are two topologically protected chiral surface states propagating in the same direction, the one-way propagation nature of the two surface states is independent of the shape of the interface as well as the local disorder. The channel sandwiched by the two phononic crystals can have sharp corners which do not induce the state switching of the two states, inside the channel the particle can be pulled towards the source irrespective of its size and material, following a trajectory confined by the channel. Although a similar idea of optical pulling using the photonic chiral surface waves [29] has been proposed recently, the acoustic pulling using the acoustic chiral surface waves has its unique advantage that it can be used to pull much larger particles with much stronger forces. Here, by rigorously calculating the acoustic forces using the closed-surface integral of the time-averaged radiation stress tensor [11-13], and quantitatively analyzing them with the response theory [30], we show that once the surface mode with the relatively small Bloch wave vector is excited, the longitudinal (along wave propagating direction) acoustic force acting on the particle located inside the channel is mostly negative. Moreover, the averaged longitudinal force for the particle moving over a lattice constant along the wave propagating direction is always negative. The negative nature of the averaged longitudinal force is guaranteed by the Bloch wave vector difference between the two surface modes. Therefore, for a particle made of any material, possessing any size, it can always be pulled backward along the interface waveguide by the surface waves due to the net pulling force.

## II. Chiral surface states

The chiral surface states in classical systems have been predicted [31-33] and realized [34,35] in 2-dimensional magneto-optical photonic crystals by breaking time-reversal symmetry in analogy to the quantum Hall effect in electronic systems. The acoustic counterparts were proposed [36-38] in phononic crystals using circulating fluid coatings, where the circulating flows break the time-reversal symmetry. This acoustic chiral edge transport was observed in experiment very recently [39].

Following the idea of Ref. [36], we consider the phononic crystal consisting of spinning cylinders with the spinning angular frequency $\Omega$ arranged into a square lattice (lattice constant $a$) in $xy$ plane embedded in water with mass density $\rho_0$ and sound speed $c_0$. When the cylinders are static, the mass density and sound speed inside are $\rho$ and $c$, respectively. A thin, sound permeable and unspinning shell is coated on each cylinder to avoid the direct contact between the cylinder and water, so that the spinning will not drive the water to move. Taking both the Doppler effect and Coriolis force into account, the $n$th order Mie coefficient of the spinning cylinder for a time-harmonic ($e^{-i\omega t}$) incident acoustic wave is given by [40, 41]

$$D_n = -\frac{\lambda_n \rho_0 R_n J_n(k_0 r_c) - k_0 \rho J_n(\lambda_n r_c) J_n'(k_0 r_c)}{\lambda_n \rho_0 R_n H_n^{(1)}(k_0 r_c) - k_0 \rho J_n(\lambda_n r_c) H_n^{(1)}{'}(k_0 r_c)}, \quad (1)$$

where $J_n$ and $H_n^{(1)}$ are the Bessel function and Hankel function of the first kind, $r_c$ is the radius of the cylinder, $k_0 = \omega / c_0$ is the wavenumber in water, $\lambda_n = \sqrt{-(M^2 + 4\Omega^2)}/c$ with $M = -i(\omega - n\Omega)$ and $c$ being the sound speed inside a static cylinder, and the auxiliary function is expressed as

$$R_n = \omega^2 \frac{(2\Omega^2 - M^2) J_n'(\lambda_n r_c) - 3in M\Omega J_n(\lambda_n r_c)/(\lambda_n r_c)}{(4\Omega^2 + M^2)(\Omega^2 + M^2)}. \quad (2)$$

When $\Omega = 0$, Eq. (1) is reduced to the Mie coefficient of a static cylinder. The band dispersion for the lattice structure containing the spinning cylinders can be obtained by using the multiple scattering method and applying the periodic boundary conditions [42, 43]. When the cylinders are static, a two-fold degeneracy at M point is enforced by the $C_{4v}$ and time-reversal symmetry, as shown in Fig. 1(a). As the cylinders are spinning, the time-reversal symmetry is broken which lifts the degeneracy and opens a topologically nontrivial band gap with gap Chern number 1 (-1) for anticlockwise (clockwise) spinning cylinders. For example, for radius $r_c = 0.16a$ and spinning frequency $\Omega = 1.678 \frac{c_0}{a}$, there is a complete band gap ranging from $0.581 \frac{c_0}{a}$ to $0.628 \frac{c_0}{a}$, see Fig. 1(b). Let us consider the interface between the two phononic crystals composed of oppositely spinning cylinders with the same spinning speed. Since the difference of the gap Chern numbers of the two sides is 2, there will exist two topologically protected chiral interface states propagating along the same direction at any frequency within the bulk band gap.

The dispersion relations of surface states can be calculated using the multiple scattering method in conjunction with supercell calculations [44]. In calculations, the supercell contains one layer along *x* direction and 18 layers along *y* direction. In order to form a closed system with purely real eigen-frequencies, all the boundaries of the supercell are connected according to periodic boundary conditions. Cylinders in the upper 9 layers have spinning frequency $\Omega$, while cylinders in the lower 9 layers have spinning frequency $-\Omega$. In addition, the upper 9 and lower 9 layers are further separated by a water channel of width $0.5a$ (therefore the dimension of the supercell is $a \times 19a$), as shown in Fig. 2(a). The band structures of the bulk (black dotted lines) and surface (red and blue dotted lines) states are shown in Fig. 2(a). We can see that there are four surface states at each frequency ranging from $0.581\frac{c_0}{a}$ to $0.628\frac{c_0}{a}$ corresponding to the band gap in Fig. 1(b). Two are propagating on the top/bottom edge (blue dotted lines) of the supercell and two are propagating on the middle (red dotted lines) interface of the supercell. At frequency $f = 0.604\frac{c_0}{a}$, the four eigen pressure field patterns corresponding to the four surface modes are shown in Fig. 2 (b). It is clearly seen that modes A and D are even while modes B and C are odd under the mirror reflection about the central line of the supercell (shown by the black dashed lines in Fig 2(b)). From the dispersion relations, we also see that both modes B and D have positive group velocities indicating that the corresponding surface waves are propagating along the positive-*x* direction.

To show that the surface states are backscattering immuned against local disorders, we did full wave simulations of the surface waves propagating on the interface and being scattered by a particle using multiple scattering technique. In simulations, 50 supercells in total are used, and an unspinning cylindrical particle is arbitrarily inside the water channel between the upper and lower phononic crystals, see Fig 2(c). To prevent the particle from entering the phononic crystals, we assume there are two hard walls (extreme thin and permeable to acoustic wave) placed at $y = \pm 0.575a$, respectively. Elastic collision occurs when the particle hits the hard walls. Therefore, there is a $1.15a \times 50a$ water gap between the two phononic crystals. Two line sources with opposite initial phases are placed at $y = \pm 0.2a$, respectively, to excite the odd mode B, We can see that the surface wave can only propagate rightward indicating that the surface wave can only transport unidirectionally. Before the wave being scattered by the particle, the pressure field distribution is odd about the middle line of the water gap since only mode B is excited. After the surface wave is scattered by the particle, the odd mirror symmetry of the field distribution is broken because a portion of mode B is scattered into the even mode D. However, the field distribution on the left hand side of the particle retains the odd symmetry indicating the surface waves are not backward scattered.

### III. Acoustic force calculation

For a time-harmonic incident acoustic wave, the time-averaged acoustic force acting on a particle submerged in water can be evaluated by integrating the stress tensor $\vec{T}$ over a surface enclosing the particle [11-13],

$$\mathbf{F} = \int_S \vec{T} \cdot d\mathbf{S}, \quad (3)$$

where the stress tensor is expressed as

$$\vec{T} = (\frac{\rho_0 \mathbf{u} \cdot \mathbf{u}^*}{4} - \frac{|p|^2}{4 c_0^2 \rho_0}) \vec{I} - \frac{\rho_0 \mathbf{u} \mathbf{u}^*}{2}, \quad (4)$$

with $p$ and $\mathbf{u}$ being the pressure and velocity field, and $\vec{I}$ being a 2 by 2 identity matrix. Using the idea of Lorenz-Mie theory [45-49], the integral can be greatly simplified when the closed surface is a circle. Then the x and y components of the acoustic force can be obtained as (see details about the derivation in the appendix)

$$F_x = \frac{p_0^2}{\rho_0 c_0^2 k_0} \text{Im} \sum_{n=-\infty}^{\infty} (2 b_n b_{n+1}^* + q_n b_{n+1}^* + b_n q_{n+1}^*),$$

$$F_y = -\frac{p_0^2}{\rho_0 c_0^2 k_0} \text{Re} \sum_{n=-\infty}^{\infty} (2 b_n b_{n+1}^* + q_n b_{n+1}^* + b_n q_{n+1}^*), \quad (5)$$

where, $p_0$ is the amplitude of the line source, $b_n$ is the scattering coefficient of the particle which is obtained through the multiple scattering calculations, and $q_n = b_n / D_n$ with $D_n$ being the Mie coefficient of the particle.

For a cylindrical particle with radius $r_s = 0.35a$, we calculated the acoustic forces acting on the particle as a function of the location $(x_s, y_s)$ of the particle confined in a rectangular region centered on the *x* axis (see the dashed rectangle in Fig. 2(c)), and the result is shown in Fig. 3(a). The rectangular region has a length *a* and a width 0.45*a*. Because the system preserves the translational symmetry, the acoustic force acting on the particle will repeat when the particle moves a lattice constant *a* along *x* direction. In Fig. 3(a), the arrows represent the acoustic force vectors and their lengths denote the force magnitudes. It can be seen that most of the acoustic forces point leftward. For each $y_s$, the spatially averaged longitudinal forces over a lattice constant $\bar{F}_x(y_s) = \frac{1}{a}\int_0^a F_x(x_s, y_s)dx_s$ are calculated and shown in Fig. 3(b) by the black squares. We can see that $\bar{F}_x(y_s)$ for every $y_s$ is negative. Thus although the particle may be subject to pushing forces at some locations, as a whole it will be accelerated leftward when it moves freely inside the gap. Even if the radius and the material component of the particle are changed, $\bar{F}_x(y_s)$ are still negative for all $y_s$, as shown by the curves of red disks and blue triangles in Fig. 3(b). In the following, we will show that so long as the incident surface wave is purely composed of mode B, $\bar{F}_x(y_s)$ is negative irrespective of the particle's size and material.

### IV. Interpretation using the response theory

The response theory was proposed previously to explain the optical force in multi-port photonic systems [30]. We apply this response theory to analyze the acoustic force in our system, as optical and acoustic forces can be described by the same mathematics.

According to the response theory adapted to our system, the longitudinal acoustic force acting on the particle fulfills

$$F_x(x_s, y_s) \propto I_B \frac{\partial \phi_B}{\partial x_s} + I_D \frac{\partial \phi_D}{\partial x_s}, (6)$$

where $I_B(I_D)$ and $\phi_B(\phi_D)$ are the intensity and phase of mode B (D) component in the scattered wave, and $(x_s, y_s)$ is the location of the particle. Taking mode B as an example, the pressure field can be expressed as $u(x,y)e^{ik_{B,0}x}$, where $u(x,y) = u(x+a, y)$ is the periodic part of the Bloch function and $k_{B,0}$ is the Bloch vector in the first Brillouin zone ($-\frac{\pi}{a} \leq k_{B,0} \leq \frac{\pi}{a}$). According to periodicity of the Bloch functions, mode B actually consists of all surface wave components which are plane waves in the $x$ direction with discrete longitudinal wave vectors $k_{B,n} = k_{B,0} + n\frac{2\pi}{a}, n \in Z$. Here we express the pressure field of $n$th surface wave component of mode B as $g(k_{B,n}, y)e^{ik_{B,n}x}$. The amplitude $|g(k,y)|$ and phase $\arg[g(k,y)]$ of the wave component with wave vector $k$ at vertical coordinate $y$ can be evaluated using the Fourier transform of the pressure field $p(x,y)$,

$$g(k,y) = \frac{1}{x_2 - x_1} \int_{x_1}^{x_2} p(x,y) e^{-ikx} dx, (7)$$

where $x_1$ and $x_2$ are the starting and ending points for the Fourier transform. The intensity of the surface wave component is proportional to $|g(k,y)|^2$ at any given $y$.

In Fig. 4, we plotted $|g(k,y)|$ at different vertical coordinates $y$ for both modes B and D. As expected, the discrete peaks of $|g(k,y)|$ are spaced by the reciprocal lattice vector $\frac{2\pi}{a}$, as shown in the insets in Fig. 4. For mode B, $|g(k,0)|$ are equal to zero due to the odd parity of the field distribution. It is worth noticing that both modes B and D are dominated by their zeroth plane wave components, although some contributions from higher order components arise for large $y$ due to the appearance of evanescent waves close to the spinning cylinders. Therefore, we can approximate modes B and D reasonably well by their zeroth components and use the two zeroth waves to study the acoustic force according to the response theory. As the intensities of the zeroth waves of modes B and D are proportional to $|g(k_{\alpha,0},y)|^2$ $(\alpha = B,D)$, respectively, the longitudinal force at position $(x_s, y)$ has the following expression under the zero order approximation

$$F_x(x_s,y) \propto \frac{|g_s(k_{B,0},y)|^2}{|g_i(k_{B,0},y)|^2}\frac{\partial}{\partial x_s}\arg[g_s(k_{B,0},y)] + (1-\frac{|g_s(k_{B,0},y)|^2}{|g_i(k_{B,0},y)|^2})\frac{\partial}{\partial x_s}\arg[g_s(k_{D,0},y)], \quad (8)$$

where $g_i$ and $g_s$ are the Fourier transforms of the incident and scattered fields respectively.

The phase and amplitude of the incident field acting on the particle will determine the amplitude and phase of the scattered field. If mode B is excited, the incident field intensity obeys even symmetry about the central vertical line of the supercell, see the inset of Fig. 5(a). Thus the intensities of modes B and D of the scattered wave are also of even symmetry about the line $x_s = 0$, as shown in Fig. 5(a). When the particle moves a lattice constant along $x$ direction, the phase of the incident wave will increase by $k_{B,0}a$. The particle is the source of the scattered waves. When the particle moves a lattice constant, the scattered waves decrease a phase of $k_{B,0}a$ and $k_{D,0}a$ relative to the incident wave, respectively. Therefore, the total phases of modes B and D of the scattered wave increase by 0 and $(k_{B,0}-k_{D,0})a$, respectively. As a result, $\phi_B$ is oscillating while $\phi_D$ decreases monotonically when $(k_{B,0}-k_{D,0})a < 0$ as the particle moves along $x$ direction, as shown in Fig. 5(b). According to Eq. (8), the oscillating of $\phi_B \sim x_s$ will contribute no net force over a lattice constant, while the monotonous decreasing of $\phi_D \sim x_s$ will lead to a negative force component. Therefore, the net optical force along $x$ direction is always negative.

In Fig. 5 (c) and (d), we compared the acoustic forces calculated by the rigorous expressions, Eq. (5), (lines) with the results calculated by the response theory, Eq. (8), (circles) for the particle moving along $x$ direction at two different vertical coordinates. The results obtained by the two calculation methods show a very good agreement, showing that the response theory works well. The slight discrepancy between the two approaches is due to the neglect of contributions from the higher order plane wave components of modes B and D.

**VI. Conclusion**

In summary, using a pair of chiral surface waves supported on the interface between two phononic crystals with broken time-reversal symmetry, we can achieve a new type of acoustic pulling. This acoustic pulling mechanism has advantages that are absent in traditional optical pulling schemes, such as a long pulling distance, flexible pulling trajectory and is independent of the particle's size and material. The band structures, transporting and scattering properties of the chiral surface waves have been studied using the multiple scattering technique. The acoustic forces acting on a particle changing with the particle's location inside the interface waveguide have been calculated rigorously using the Lorenz-Mie formula and analyzed according to the response theory in conjunction with the symmetry analysis. When the incident chiral surface mode has a smaller Bloch wave vector, the excitation of the other chiral surface mode due to the scattering by the particle will result in an averaged pulling force acting on the particle. The particle can be pulled towards the source irrespective of its size and material. Owing to the backscattering immunity of the chiral surface waves against local disorders, the channel between the two phononic crystals does not need to be straight line. If the defect on the interface does not break the $C_{4v}$ symmetry [30], such as a corner with 90º bending angle, mode transform cannot occur before the wave is scattered by the particle. So the particle can be pulled continuously in a channel that possesses an arbitrary number of right-angled bends. As there is no restriction on the length of the channel, a long pulling distance can be easily achieved. Our work shows that topological sound waves can be used to control particles more proficiently than ordinary acoustic waves.

**Acknowledgments**: We thank Prof. Zhao-Qing Zhang for constructive suggestions. This work is supported by National Natural Science Foundation of China (NSFC) through No. 11904237 and

Hong Kong Research Grants Council through grant No. AoE/P-02/12 and 16303119.

**Appendix: Lorenz-Mie theory for acoustic force**

Choosing the center of the cylindrical particle as the origin, the pressure field at $\mathbf{r} = (r, \phi)$ according to the Mie theory [49] can be expressed as

$$p(\mathbf{r}) = p_0 \sum_{n=-\infty}^{\infty} [q_n J_n(k_0 r) + b_n H_n^{(1)}(k_0 r)] e^{in\phi}, \quad (A1)$$

where $b_n$ is the scattering coefficient of the particle, and $q_n = b_n / D_n$ with $D_n$ being the Mie coefficient of the particle. The velocity field is calculated according to

$$\begin{aligned} \mathbf{u} &= -\frac{i}{\omega \rho_0} \nabla p \\ &= -\frac{ip_0}{\omega \rho_0} \sum_{n=-\infty}^{\infty} \{[q_n k_0 J_n'(k_0 r) + b_n k_0 H_n^{(1)'}(k_0 r)]\hat{e}_r + \frac{in}{r}[q_n J_n(k_0 r) + b_n H_n(k_0 r)]\hat{e}_\phi\} e^{in\phi}. \end{aligned} \quad (A2)$$

The acoustic force is obtained by substituting Eqs. (A1) and (A2) into Eq. (3). Due to the law of momentum conservation, the closed surface in the integral can be arbitrary. For the sake of simplicity, we choose the closed surface as a circle with infinite radius centered at the origin. At the infinity $r \to \infty$, using the asymptotical formulae of Bessel and Hankel functions,

$$\begin{aligned} J_n(x) &\approx \sqrt{\frac{2}{\pi x}} \cos(x - \frac{n\pi}{2} - \frac{\pi}{4}), \quad H_n(x) \approx \sqrt{\frac{2}{\pi x}} e^{i(x - \frac{n\pi}{2} - \frac{\pi}{4})}, \\ J_n'(x) &\approx -\sqrt{\frac{2}{\pi x}} \sin(x - \frac{n\pi}{2} - \frac{\pi}{4}), \quad H_n'(x) \approx i\sqrt{\frac{2}{\pi x}} e^{i(x - \frac{n\pi}{2} - \frac{\pi}{4})}, \end{aligned} \quad (A3)$$

and ignoring the high order terms of $1/r$, the pressure and velocity fields are reduced to

$$\begin{aligned} p(\mathbf{r}) &= p_0 \sqrt{\frac{2}{\pi x}} \sum_{n=-\infty}^{\infty} (q_n \cos y_n + b_n e^{iy_n}) e^{in\phi}, \\ \mathbf{u}(\mathbf{r}) &= -\frac{ip_0}{\omega \rho_0} \sqrt{\frac{2}{\pi x}} \sum_{n=-\infty}^{\infty} (-q_n \sin y_n + ib_n e^{iy_n}) e^{in\phi}, \end{aligned} \quad (A4)$$

where $x = kr, y_n = x - n\pi/2 - \pi/4$. Substituting Eq. (A4) into Eq. (3),

$$f_x = \hat{e}_x \cdot \int_S \ddot{\mathbf{T}} \cdot d\mathbf{S} = -\frac{1}{4}\int_0^{2\pi}(\rho_0|\mathbf{u}|^2 + \frac{1}{\kappa_0}|p|^2)r\cos\phi d\phi$$

$$= -\frac{|p_0|^2}{2\rho_0 c_0^2 k_0}\{\sum_n \begin{array}{l}[(-q_n\sin y_n + ib_n e^{iy_n})(-q_{n+1}^*\sin y_{n+1} - ib_{n+1}^* e^{-iy_{n+1}}) \\ +(-q_n\sin y_n + ib_n e^{iy_n})(-q_{n-1}^*\sin y_{n-1} - ib_{n-1}^* e^{-iy_{n-1}})]\end{array} \quad (A5)$$

$$+\sum_n \begin{array}{l}[(q_n\cos y_n + b_n e^{iy_n})(q_{n+1}^*\cos y_{n+1} + b_{n+1}^* e^{-iy_{n+1}}) \\ +(q_n\cos y_n + b_n e^{iy_n})(q_{n-1}^*\cos y_{n-1} + b_{n-1}^* e^{-iy_{n-1}})]\end{array}\}.$$

Note that

$$\sin y_{n+1} = \sin(y_n - \frac{\pi}{2}) = -\cos y_n, \cos y_{n+1} = \sin y_n,$$

$$\sin y_{n-1} = \sin(y_n + \frac{\pi}{2}) = \cos y_n, \cos y_{n-1} = -\sin y_n, \quad (A6)$$

$$e^{-iy_{n+1}} = e^{-iy_n}e^{i\frac{\pi}{2}} = ie^{-iy_n}, e^{-iy_{n-1}} = e^{-iy_n}e^{-i\frac{\pi}{2}} = -ie^{iy_n},$$

acoustic force along x direction becomes

$$f_x = -\frac{|p_0|^2}{2\rho_0 c_0^2 k_0}\sum_{n=-\infty}^{\infty}\{ib_n q_{n+1}^* + 2ib_n b_{n+1}^* - ib_n q_{n-1}^* - 2ib_n b_{n-1}^* + iq_n b_{n+1}^* - iq_n b_{n-1}^*\}$$

$$= \frac{|p_0|^2}{\rho_0 c_0^2 k_0}\text{Im}\sum_{n=-\infty}^{\infty}(2b_n b_{n+1}^* + b_n q_{n+1}^* + q_n b_{n+1}^*). \quad (A7)$$

Similarly, the force along y direction is

$$f_y = -\frac{|p_0|^2}{\rho_0 c_0^2 k_0}\text{Re}\sum_{n=-\infty}^{\infty}(2b_n b_{n+1}^* + p_n b_{n+1}^* + b_n p_{n+1}^*). \quad (A8)$$

**References:**

1. A. Eller, J. Acoust. Soc. Am. 43, 170 (1968).

2. J. Wu, J. Acoust. Soc. Am. 89, 2140 (1991).

3. D. Baresch, J.-L. Thomas, and R. Marchiano, Phys. Rev. Lett. 024301 (2016).

4. A. Anhauser, R. Wunenburger, and E. Barsselet, Phys. Rev. Lett. 109, 034301 (2012).

5. S. Santesson and S. Bilsson, Anal. Bioanal. Chem. 378, 1704 (2004).6

6. R. Tuckermann, L. Puskar, M. Zavabeti, R. Sekine, and D. McNaughton, Anal. Bioanal. Chem.

394, 1433 (2009).

7. J. Shi, D. Ahmed, X. Mao, S.-C. S. Lin, A. Lawit, and T. J. Huang, Lab Chip 9, 2890 (2009).

8. X. Ding et al., PNAS 109, 11105 (2012).

9. F. Guo et al., PNAS 113, 1522 (2016).

10. W. J. Xie, C. D. Cao, Y. J. Lu, Z. Y. Hong, and B. Wei, Appl. Phys. Lett. 89, 214102 (2006).

11. P. L. Marston, J. Acoust. Soc. Am. 120, 3518 (2006).

12. P. L. Marston, J. Acoust. Soc. Am. 122, 3162 (2007).

13. P. L. Marston, J. Acoust. Soc. Am. 125, 3539 (2009).

14. L. Zhang et al., Phys. Rev. E 84, 035601(R) (2011).

15. L. Zhang et al., J. Acoust. Soc. Am. 131, EL329 (2012).

16. S. Xu, C. Qiu, and Z. Liu, Europhys. Lett. 99, 44003 (2012).

17. C. E. M. Demore et al., Phys. Rev. Lett. 112, 174302 (2014).

18. F. G. Mitri, J. Appl. Phys. 117, 094903 (2015).

19. E. H. Trinh, Rev. Sci. Instrum. 56, 2059 (1985).

20. A. L. Yarin, M. Pfaffenlehner, and C. Tropea, J. Fluid Mech. 356, 65 (1998).

21. E. H. Brandt, Nature 413, 474 (2001).

22. A. Marzo, S. A. Seah, B. W. Drinkwater, D. R. Sahoo, B. Long, and S. Subramanian, Nat. Coummun. 6, 8661 (2015).

23. H. M. Hertz, J. Appl. Phys. 78, 4845 (1995).

24. J. Lee, S.-Y. Teh, A. Lee, H. H. Kim, C. Lee, and K. K. Shung, Appl. Phys. Lett. 95, 073701 (2009).

25. D. J. Collins, C. Devendran, Z, Ma, J. W. Ng, A. Neild, and Y. Ai, Sci. Adv. 2, e1600089 (2016).

26. S. C. Takatori, R. D. Dier, J. Vermant, and J. F. Brady, Nat. Commun. 7, 10694 (2016).

27. F. Guo, P. Li, J. B. French, Z. Mao, H. Zhao, S. Li, N. Nama, J. R. Fick, S. J. Benkovic, and T. J. Huang, Proc. Natl. Acad. Sci. 112, 43 (2015).

28. A. Marzo and B. W. Drinkwater, Proc. Natl. Acad. Sci. 116, 84 (2019).

29. D. Wang, C.-W. Qiu, P. T. Rakich, and Z. Wang, CLEO (2015).

30. P. T. Rakich, M. A. Popovic, and Z. Wang, Opt. Express **17**, 18116 (2009).

31. F. D. M. Haldane and S. Raghu, Phys. Rev. Lett. 100, 013904 (2008).

32. S. Raghu and F. D. M. Haldane, Phys. Rev. A 78, 033834 (2008).

33. Z. Wang, Y. D. Chong, J. D. Joannopoulos, and M. Soljacic, Phys. Rev. Lett. 100, 013905 (2008).

34. Z. Wang, Y. Chong, J. D. Joannopoulos, and M. Soljacic, Nature **461**, 772 (2009).

35. Y. Poo, R.-X. Wu, Z. Lin, Y. Yang, and C. T. Chan, Phys. Rev. Lett. **106**, 093903 (2011).

36. Z. Yang, F. Gao, X. Shi, X. Lin, Z. Gao, Y. Chong, B. Zhang, Phys. Rev. Lett. 114, 114301 (2015).

37. A. B. Khanikaev, R. Fleury, H. Mousavi, and A. Alu, Nat. Commun. 6, 8260 (2015)

38. X. Ni, C. He, X.-C. Sun, X. Liu, M.-H. Lu, L. Feng, and Y.-F. Chen, New. J. Phys. 17, 053016 (2015).

39. Y. Ding, Y. Peng, Y. Zhu, X. Fan, J. Yang, B. Liang, X. Zhu, X. Wan, and J. Cheng, Phys. Rev. Lett. 122, 014302 (2019).

40. D. Censor and J. Aboudi, Journal of Sound Vibration 19, 437 (1971).

41. D. Zhao, Y.-T. Wang, K.-H. Fung, Z.-Q. Zhang, and C. T. Chan, Phys. Rev. B 101, 054107 (2020).

42. J. S. Faulkner, Phys. Rev. B 19, 6186 (1979).

43. S. K. Chin, N. A. Nicorovici, and R. C. McPhedran, Phys. Rev. E 49, 4590 (1994).

44. X. Zhang, L.-M. Li, Z.-Q. Zhang, and C. T. Chan, Phys. Rev. B 63, 125114 (2000).

45. G. Gouesbet, J. Quant. Spectrosc. Radiat. Transf. 110, 1223 (2009).

46. G. Gouesbet and G. Grehan, Generalized Lorenz-Mie Theories (Springer, Berlin, 2011).

47. G. Gouesbet and J. A. Lock, Appl. Opt. 52, 897 (2013).

48. N. Wang, J. Chen, S. Liu, and Z. Lin, Phys. Rev. A 87, 063812 (2013).

49. C. F. Bohren and D. R. Huffman, absorption and Scattering of Light by Small particles (John Wiley and Sons, New York, 1983).

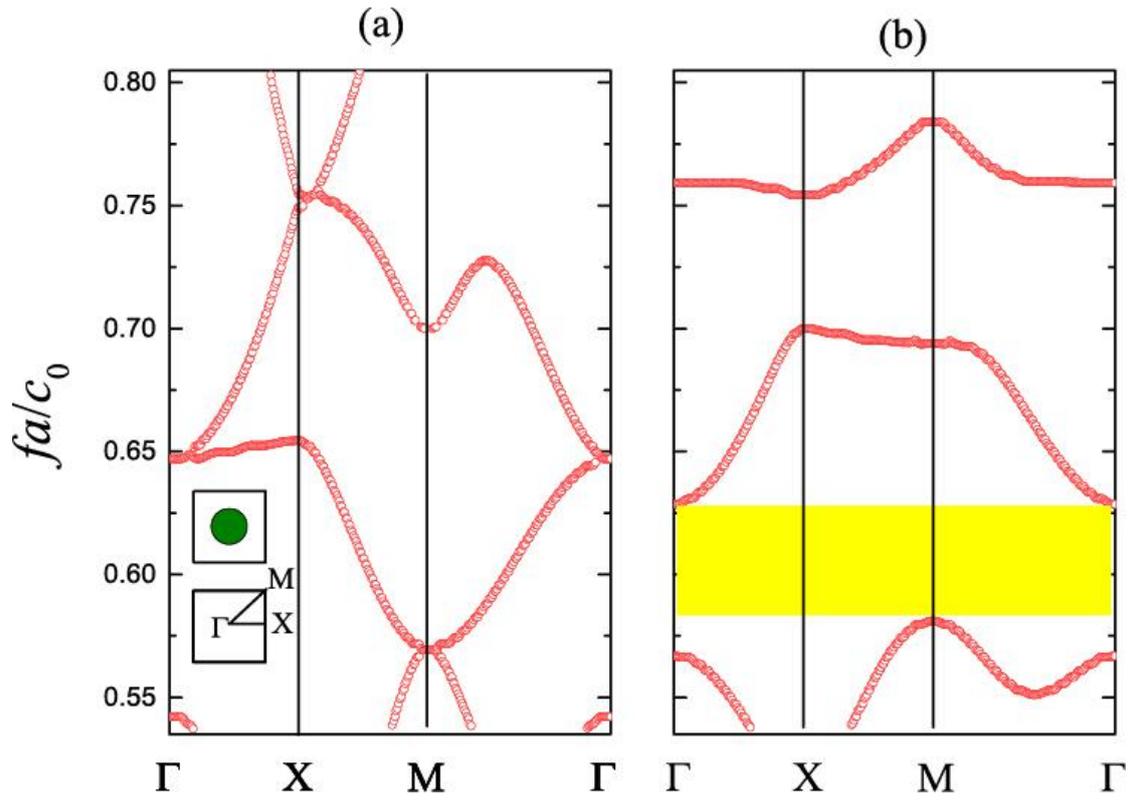

Fig. 1. Band structures of phononic crystals composed of (a) static cylinders and (b) spinning cylinders arranged into a square lattice embedded in water. The parameters of the water and static cylinder are $\rho_0 = 10^3 kg/m^3, c_0 = 1489 m/s, \rho = 1.3 \times 10^3 kg/m^3, c = 400 m/s$, the radius of the cylinder is $r_c = 0.16a$, and the spinning circular frequency is $\Omega = 1.678a/c_0$.

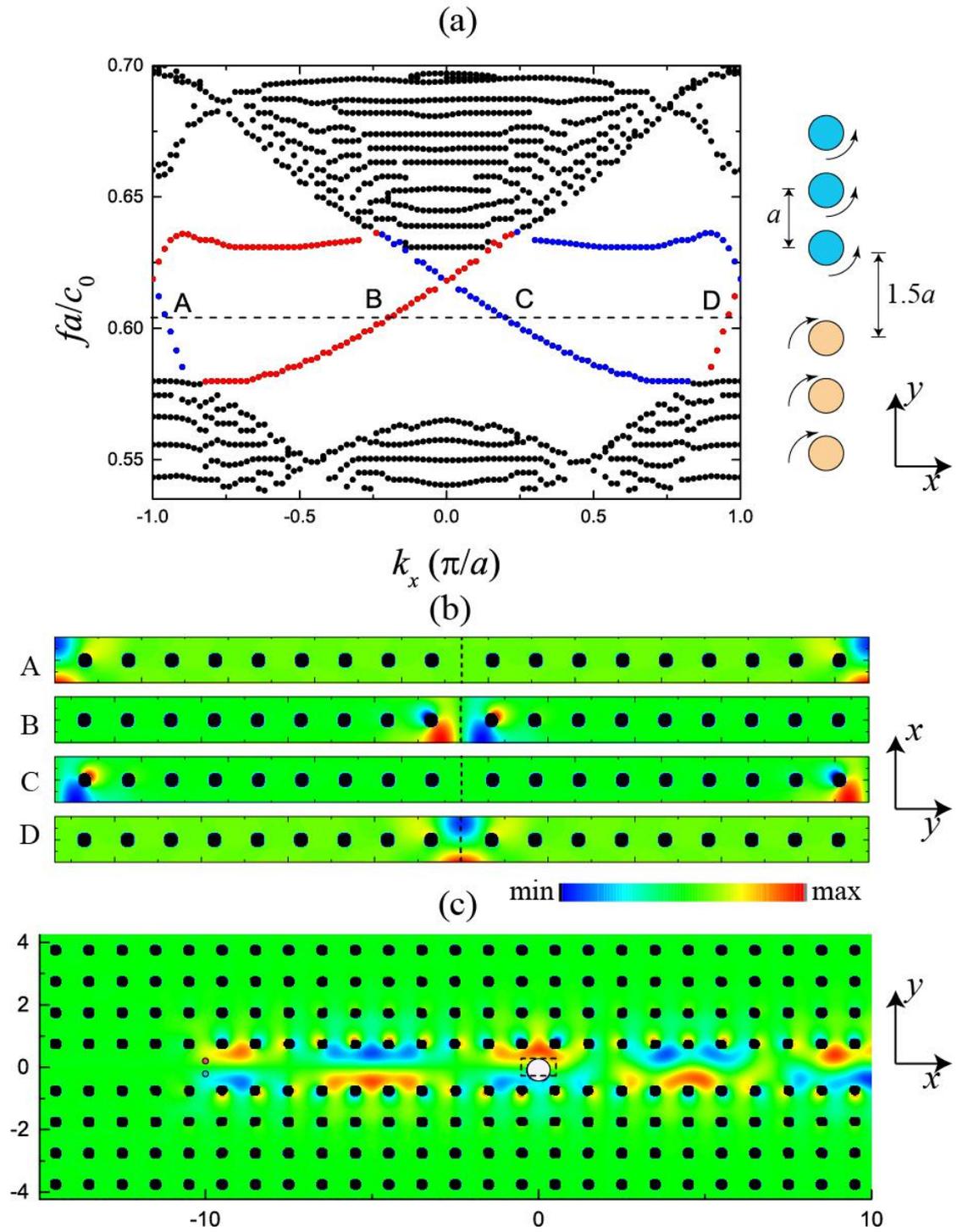

Fig. 2. (a) Bulk (black) and surface (red and blue) state dispersion relations calculated using multiple scattering method in conjunction with a supercell calculation. (b) Pressure field patterns for the four surface modes. (c) Full wave simulations for the one-way surface wave being scattered by a cylindrical particle. The spinning cylinders are shown by the black disks in (b) and (c). In (c), the scatter is represented by the white disk, and the two line sources (red and blue dots) $p_0 H_0^{(1)}(k_0 r_1)$ and $-p_0 H_0^{(1)}(k_0 r_2)$ are located at $\mathbf{r}_1 = (-10, 0.2)$ and $\mathbf{r}_2 = (-10, -0.2)$,

respectively.

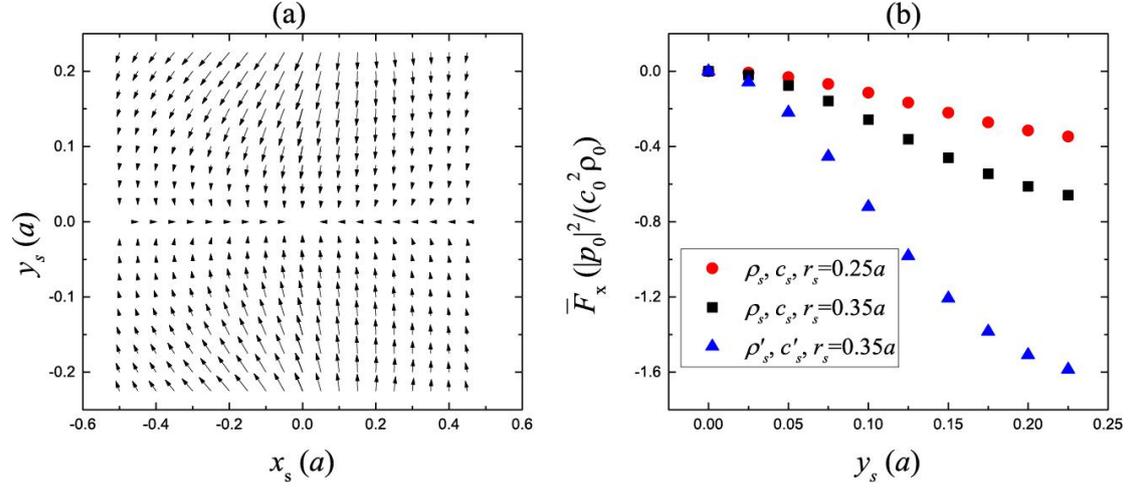

Fig. 3. (a) The acoustic forces acting on the particle as a function of the location $(x_s, y_s)$ of the particle within a rectangular region as marked in Fig. 2(c). The parameters of the cylinder are $r_s = 0.35a, \rho_s = 7800 kg/m^3, c_s = 6010 m/s$. (b) The spatially averaged longitudinal acoustic force $\bar{F}_x$ as functions of $y_s$ for different particles. For the red black symbol lines, the mass density and sound speed of the particle are $\rho_s = 7800 kg/m^3, c_s = 6010 m/s$, while for the blue symbol lines, the mass density and sound speed of the particle are $\rho'_s = 1.3 kg/m^3, c'_s = 340 m/s$.

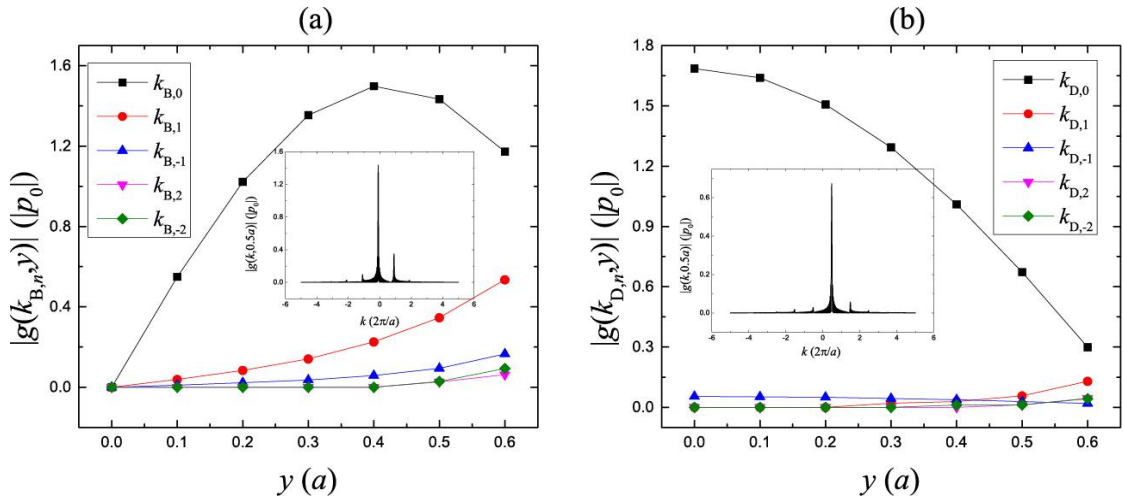

Fig. 4. (a) $|g(k_{B,n}, y)|$ and (b) $|g(k_{D,n}, y)|$ as functions of the vertical coordinate $y$. The insets show $|g(k, y)|$ as functions of plane wave vector $k$ at $y = 0.5a$.

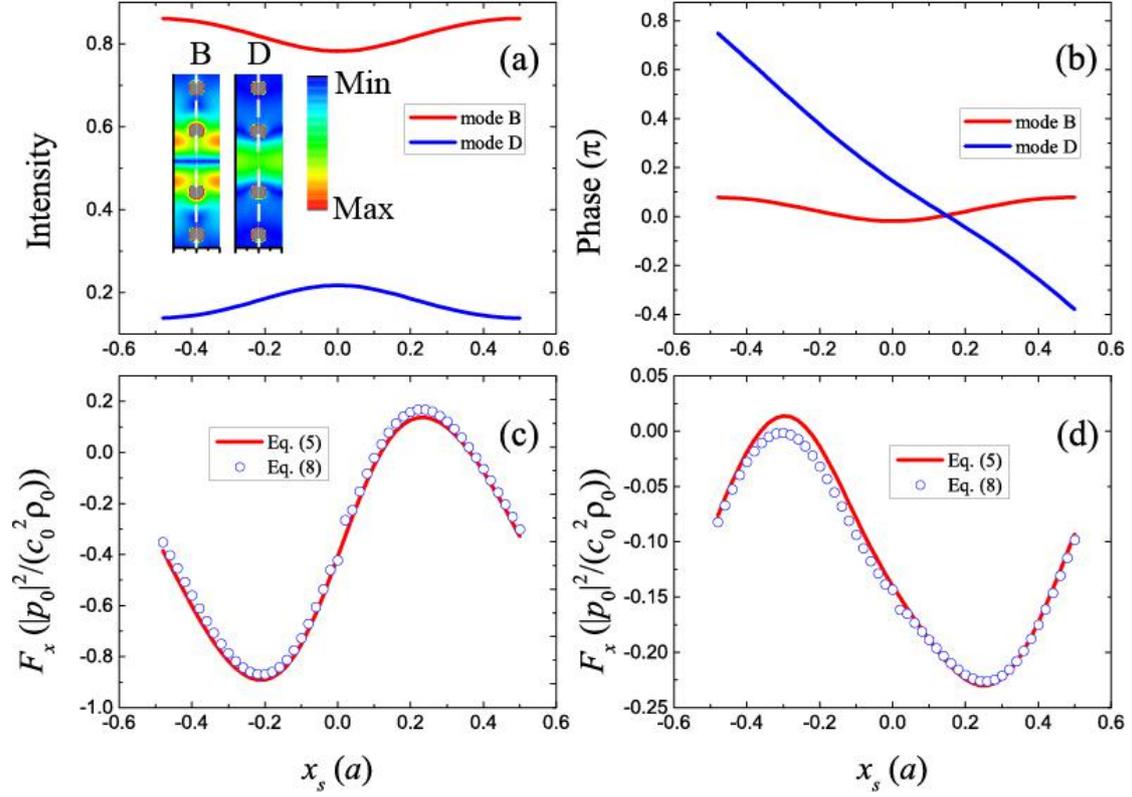

Fig. 5. (a) and (b) the intensities $I_B, I_D$, and relative phases $\phi_B, \phi_D$ of modes B and D of zero order in the scattered wave which are normalized by the intensity of the incident wave as functions of the particle location $x_s$. The vertical coordinate of the particle is $y_s = 0.3a$. The insect in (a) shows the intensity distributions of eigenmodes B and D. (c) and (d) the longitudinal acoustic forces $F_x$ acting on the particle as functions of the particle location $x_s$ which are calculated using the Lorenz-Mie formula Eq. (5) (red lines) and response theory Eq. (8) (blue circles). The vertical coordinates of the particle in (c) and (d) are $y_s = 0.3a$ and $y_s = 0.1a$ respectively. The parameters of the particle are $r_s = 0.25a$, $\rho_s = 7800 kg/m^3$, $c_s = 6010 m/s$.